\documentclass[onecolumn]{IEEEtran}

\usepackage{times}
\usepackage{epsfig}
\usepackage{graphicx}
\usepackage{amsmath}
\usepackage{amssymb}

\begin{document}

\title{Market Fragility, Systemic Risk, and Ricci Curvature}

\author{Romeil Sandhu, Tryphon Georgiou, Allen Tannenbaum\\
{\tt\small romeil.sandhu@stonybrook.edu, tryphon@umn.edu, allen.tannenbaum@stonybrook.edu}
}
\maketitle
\thispagestyle{empty}

\begin{abstract}
Measuring systemic risk or fragility of financial systems is a ubiquitous task of fundamental importance in analyzing market efficiency, portfolio allocation, and containment of financial contagions.   Recent attempts have shown that representing such systems as a weighted graph characterizing the complex web of interacting agents over some information flow (e.g., debt, stock returns, shareholder ownership) may provide certain keen insights. Here, we show that fragility, or the ability of system to be prone to failures in the face of random perturbations, is negatively correlated with geometric notion of Ricci curvature.  The key ingredient relating fragility and curvature is entropy.   As a proof of concept, we examine returns from a set of stocks comprising the S\&P 500 over a 15 year span to show that financial crashes are more robust compared to normal  ``business as usual'' fragile market behavior - i.e., Ricci curvature is a ``crash hallmark.''  Perhaps more importantly, this work lays the foundation of understanding of how to design systems and policy regulations in a manner that can combat financial instabilities exposed during the 2007-2008 crisis.

\end{abstract}

\section{Introduction}
\begin{figure}[hbt]
\begin{center}
\includegraphics[height=6cm]{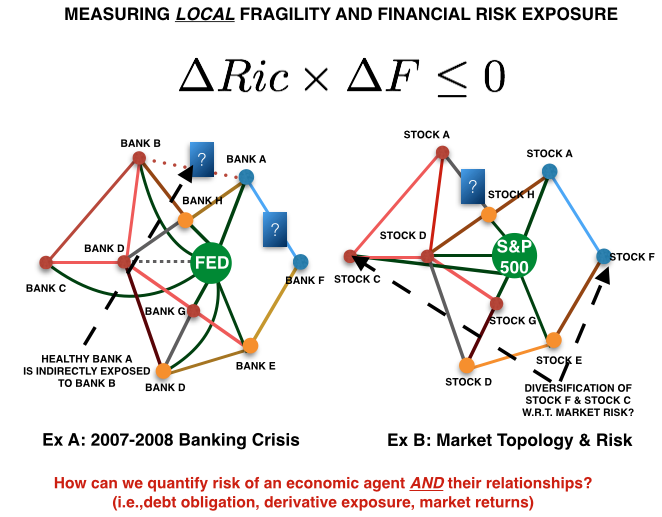}
\caption{Understanding indirect counter-party risk has gained increasing importance with the recent global financial crisis combined with the continue rise of complex financial instruments.  This paper proposes a new metric for characterizing the instability with respect agent-to-agent information in the context of a global network.  As an example, we show initial proof-of-concept analysis on analyzing market fragility from a feedback perspective resulting from well-known ``herding'' phenomena during periods of financial crisis.}
\label{fig:sweet_spot}
\end{center}
\end{figure}

The financial crisis of 2007-2008 has put much attention in understanding global-to-local fragility of financial systems \cite{schweitzer}.  A recent accepted model is one in which these interconnected systems are represented as a weighted graph whereby the nodes denote an economic agent and the edge links characterize dependencies between such agents (e.g., returns, debt, derivative exposure) \cite{bat1,haldane1,man1,onnel1}.  In turn, systemic risk can be taken as a network's inability to handle default of one or more agents resulting in cascade failures triggering the onset of a financial contagion. That is, in a given financial network, one must be able to attribute a proper measure of risk to those institutions (nodes), and more importantly their relationships (edges) that are often deemed ``too-big-to-fail.''  This seems especially true during the emergency aid providing by the U.S. Federal Reserve Bank (FED) during the 2007-2008 crisis \cite{FED}.  As seen in Figure~\ref{fig:sweet_spot}, there exists a series of indirect ``hidden'' exposures that unraveled during this time period.   Understanding fragility in the context of financial systems provides not only a measure of preventing (combating) financial contagions, but may also provide novel insights into designing downside protection (VaR) measures.

The goal of this paper is to demonstrate Ricci curvature may serve as an indicator of fragility in the context of financial networks.  Curvature, in the broad sense, is a measure by which a geometrical object deviates from being flat, and has been characterized in various ways in differential geometry \cite{DoCarmo}. Recently, this notion has been extended to more general frameworks such as networks modeled as weighted graphs \cite{Oll_markov,Ollivier,LV,Mass,Zhou}. In the present work, we examine the topology of a stock correlation networks constructed from the S\&P 500 over a 15 year span and show that curvature is a ``crash hallmark.''  While the initial analysis herein is restricted to correlation networks, one can naturally apply our analysis to information quantities that pertain to the banking ecosystem \cite{bat1,haldane1} or more complex financial instruments \cite{bat3}.  The usage of stock returns is motivated in a two-fold manner:  Firstly, we use the evolution of the stock market simply as a \emph{data platform} to primarily illustrate a fascinating connection between Ricci curvature and a well established notion of entropy (and hence fragility).  Secondly, the stock data is readily obtained and is not obfuscated due to proprietary knowledge of the institutions themselves - further data sets are currently being gathered for analysis.  This said, we first revisit several key papers that are related to the present work.

\subsection{Relation to Previous Work}
First and foremost, we note that this paper is a follow-on and complementary effort of previous work whereby we introduced Ricci curvature as a proxy for robustness in order to differentiate cancer networks \cite{cancer_arxiv}.  We now turn our attention to financial networks and note that information presented herein overlaps with existing concepts in order for this manuscript to be self-contained (and refer the reader to \cite{cancer_arxiv} for compete details).  This being said, with the emergence of network science, recent advancements have explored the concepts of fragility on weighted graphs (and the construction thereof) \cite{dem,barabasi,barabasi2,demo}.  In particular, Demetrius \cite{dem,fluctuation} formally defines ``robustness'' in terms of the rate function from the theory of large deviations and shows that this is positively correlated to entropy via the \textbf{\emph{Fluctuation Theorem}}; we will describe this in some details in Section~\ref{sec:prelim_FT}.  The key difference in our approach is that Ricci curvature serves as a proxy for robustness/fragilty \emph{at the edge level} of any weighted graph as compared to entropy, which is defined as a nodal attribute.  That is, unlike entropy which, by construction, exhibits a ``loss of information'' due to a weighted contraction of edge dependencies, Ricci curvature preserves such geometric quantities.

In the context of financial networks, various works seek to examine the fragility of interactions (edges) to better characterize market complexity \cite{bat1,bat3,bat4,Jackson,Allen} -- a complete review is beyond the scope of this short note, but we highlight several recent studies.  Recently, Battiston \cite{bat1} proposed a concept of DebtRank to analyze systemic risk due to debt obligations for general banking environments \cite{haldane1}, further expounding on the need to understand instability of complex derivatives \cite{bat3} to more tacit areas of corporate control in economic networks \cite{bat2}.  The main thrust and motivation in these works is ``feedback centrality,'' for which we introduce the concept of Ricci curvature as a feedback measure.  With respect to stock correlation networks, Mantenga \cite{man1} first illustrate the hierarchical arrangement of stocks through minimum spanning trees (MST).
This was followed by several works by Onnella \cite{onnel1, onnel2,onnel3} leveraging the concept of MST to exploit the underlying return dynamics.  In particular, the authors then show that during crash periods, the tree structure ``shrinks and tightens'' compared to normal market behavior and perhaps more interestingly, those stocks that serve as ``leafs'' of the tree correlate to diversification with respect to portfolio construction pioneered by Markowitz \cite{marko}.  This work seeks to cast existing work on correlation networks in the context of curvature, fragility, and uncertainty while also paving the way to analyze more interesting financial networks commonly seen in interbanking lending markets.

The remainder of the present paper is outlined as follows: In the next section, we first provide important preliminaries regarding the Fluctuation Theorem and $L^p$ Wasserstein distance.  Section~\ref{sec:proposed_method} motivates Ricci curvature as a fragility measure as well as a compatible notion of Ricci curvature on discrete metric spaces proposed by Ollivier \cite{Oll_markov,Ollivier}.  Then, Section~\ref{sec:results} provides information on the construction of financial networks under investigation, experimental results demonstrating the fragility of normal market behavior and robustness of market crashes, and a comparison of curvature to that of well established notions of graph robustness.  Lastly, Section \ref{sec:summary} concludes with a summary and discussion of future work.

\section{Preliminaries}
In this section, we review necessary concepts that we will need in the sequel.

\subsection{Fluctuation Theorem}
\label{sec:prelim_FT}
We first begin by providing a precise definition of robustness (and hence fragility) and relate this to entropy via the Fluctuation Theorem formulated in \cite{demo,fluctuation}.  Given a network, one can consider a random perturbation that results in a deviation of some observable. More formally, let $q_\delta (t)$ denote the
probability that the mean deviates by more than $\delta$ from the original value at some time $t$. Then, if we assume $q_\delta(t) \rightarrow 0$ as $t \rightarrow \infty,$ the relative rate that system ``relaxes'' and returns to its unperturbed state measures its fragility (e.g., longer decay rates are analogous to more fragile states) is given by
\begin{equation}
R := \lim_{t \rightarrow \infty} (-\frac{1}{t} \log q_\delta(t)).
\end{equation}
Therefore, large $R$ means not much deviation and small $R$ corresponds to large deviations.

In thermodynamics it is well-known that entropy and rate functions from large deviations are very closely related \cite{varadhan}. The Fluctuation Theorem is an expression of this fact for networks and may be expressed in terms of fragility $\Delta F = - \Delta R$ as
\begin{equation}
\Delta S_e \times \Delta F \leq 0.
\end{equation}
where $S_e$ denotes network entropy defined in \cite{demo}.  Specifically, let us consider a stochastic matrix $\phi= (\eta_{xy})$ describing a Markov chain that characterizes transition rates from state $x\rightarrow y$ with $\eta_{xy}\geq 0$ and $\sum_y \eta_{xy}=1$ (along with its invariant distribution $\pi=\pi \phi$).  Then network entropy may be defined as:
\begin{equation}
\label{eq:network_entropy}
S_e = \sum_x \pi(x) \bar{S}(x) \quad \text{with} \quad \bar{S}(x)=-\sum_y\eta_{xy}\log \eta_{xy}
\end{equation}
We note that in the above definition, the nodal entropy $\bar{S}(x)$ is a summation only over edges $y$ emanating towards adjacent vertices.  This is particularly important as it first discounts information for non-adajacent vertices; the computation for $n$-step random walks (Markov processes) with $n>1$ may become computationally expensive \cite{West}.  As we shall soon see, Ricci curvature can be formulated as a simple linear program and is not restricted to direct incidences.  Secondly, and perhaps more importantly, local (nodal) entropy ``loses information'' with respect to edge information through a weighted contraction -- \textbf{this quantification of edge fragility is precisely what Ricci curvature provides}.

\subsection{Wasserstein Distance}
We now turn our attention by recording the basic definition of the $L^{p}$-Wasserstein distance from optimal transport theory that we will need below.  For full details about the Monge-Kantorovich (optimal mass transport) problem and the associated Wasserstein distance, we refer the reader to several works on this topic\cite{Evans1989,Rachev,Villani2,Villani3,Tannenbaum2010signals, Givens}.  The $L^p$ \emph{Wasserstein distance} is defined as
\begin{equation}
W_p(\mu_1,\mu_2) := \bigg( \inf_{\mu\in\Pi(\mu_1,\mu_2)} \int\int d(x,y)^p d\mu(x,y)\bigg)^{1/p} \nonumber
\end{equation}
where $X$ is a metric measure space equipped with distance $d$, $\Pi (\mu_1,\mu_2)$ is a set of all couplings between two measures $\mu_1$ and $\mu_2$ with the same total mass and finite $p$-th moment.  More precisely, a \emph{coupling} between $\mu_1$ and $\mu_2$ is a measure $\mu$ on $X \times X$ such that
\begin{equation}
\int_y d\mu(x,y) = d\mu_1(x), \;\;\;\; \int_x d\mu(x,y) = d\mu_2(y) \nonumber
\end{equation}
In other words, the marginals of $\mu$ are $\mu_1$ and $\mu_2$.  In this paper, we only consider the cases $p=1,2$.

\section{Proposed Method}
\label{sec:proposed_method}
This section introduces Ricci curvature as a proxy for robustness and fragility.  We then provide one compatible notion of Ricci curvature on a discrete metric space (graphs) proposed by Ollivier~\cite{Ollivier}. Among a number of considerations, this coarse geometric definition is motivated in the classic Ornstein-Uhlenbeck process and placed in the context of fat-tailed distributions.

\subsection{Ricci Curvature and Fragility}
There have been a number of approaches \cite{Ollivier,Romania,Mass,Zhou,LV,McC97, tetali, Sturm, Gromov} to extending the notion of Ricci curvature to more general metric measure spaces. At this point, the exact relationship of one approach as compared to another is unclear.  In particular, we follow
\cite{Ollivier,Romania} because of connections to notions of metric entropy.  That is, if we let $(X,d,m)$ denote a geodesic space and set
\begin{eqnarray}
P(X,d,m)&\!\!\!\!\!\!:=&\!\!\!\!\!\! \{ \mu \ge 0:\!\!\!\int_X \mu \, dm =1\},\notag\\
P^*(X,d,m)&\!\!\!\!\!\! :=&\!\!\!\!\!\! \{ \mu \in P(X,d,m):\!\!\lim_{\epsilon \searrow 0} \int_{\mu \ge \epsilon}\!\!\!\!\!\! \!\!\!\mu \log \mu \, dm < \infty\} \notag\end{eqnarray}
then, Lott and Villani \cite{LV} showed that $X$ has {\em Ricci curvature bounded from below by $k$} if
for every $\mu_0, \mu_1 \in P(X),$ there exists a constant speed geodesic $\mu_t$ with respect to the Wasserstein 2-metric connecting $\mu_0$ and
$\mu_1$ such that
\begin{equation}
\label{eq:entropy.curvature}S_e(\mu_t) \ge t S_e(\mu_0) + (1-t) S_e(\mu_1)+\frac{kt(1-t)}{2} W(\mu_0, \mu_1)^2
\end{equation}
for $0\le t\le 1,$ and where we define
\begin{equation}\label{Bent}
\mathcal{H}(\mu) := \lim_{\epsilon \searrow 0} \int_{\mu \ge \epsilon} \mu \log \mu \, dm,
\mbox{ for } \mu \in P^*(X,d,m),
\end{equation}
which is the negative of the {\em Boltzmann entropy}
$S_e(\mu):=-\mathcal{H}(\mu)$.  This indicates that entropy and curvature are \textbf{\emph{positively correlated}} for which express as
\begin{equation} \label{ricci.entropy} \Delta S_e \times \Delta Ric \ge 0.\end{equation}
We note here that changes in {\em fragility}, i.e., a system that is prone to failures and unable to adapt to changes in the environment is \textbf{\emph{negatively correlated}} with entropy via the Fluctuation Theorem \cite{demo,fluctuation}, and thus with Ricci curvature:
\begin{equation} \label{ricci.robustness} \Delta F \times \Delta Ric \leq 0.\end{equation}
However, the above definition is intractable on a discrete space of graphs for which we are most interested in.  To this end, we adopt the recent notion of Ollivier-Ricci curvature motivated from coarse geometry \cite{Oll_markov,Ollivier}.

\subsection{Ollivier-Ricci Curvature}
\label{sec:or_curvature}
We employ a neat notion of a Ricci curvature \cite{Ollivier} inspired through coarse geometry.  The idea is motivated from the notion that the distance between two small (geodesic) balls is less than the distance of their centers.  In particular, if we let $(X,d)$ be metric space equipped with a family of probability measures $\{\mu_x : x \in X\}$, we define the {\em Ollivier-Ricci curvature} $\kappa(x,y)$ along the geodesic connecting nodes $x$ and $y$ via
\begin{equation} \label{OR}
W_1(\mu_x,\mu_y) = (1-\kappa(x,y))d(x,y),
\end{equation}
where $W_1$ denotes the Earth Mover's Distance (Wasserstein 1-metric) \cite{Villani2,Villani3}, and $d$ is the geodesic distance on the graph.
For the case of weighted graphs, we set
\begin{eqnarray*} d_x &=& \sum_{y} w_{xy}\\
\mu_x(y)&:=& \frac{w_{xy}}{d_x} ,\end{eqnarray*}
where $d_x$ is the sum taken over all neighbors of node $x$ and where $w_{xy}$ denotes the weight of an edge connecting node $x$ and node $y$ ($w_{xy}=0$ if $d(x,y)\geq 2$). The measure $\mu_x$ may be regarded as the distribution of a one-step random walk starting from $x$, with the weight $w_{xy}$ quantifying the strength of interaction between nodal components or the diffusivity across the corresponding link (edge). Furthermore,if $\mu_1$ and $\mu_2$ are two distributions represented by $n$ discrete points $\{x_1, \ldots, x_n\}$ having the same total mass and denote the distance $d(x,y)$ between $x,y \in X$ (for graphs, taken as the hop metric), then $W_1(\mu_1,\mu_2)$ may be defined as follows \cite{Rubner}:
\begin{equation}
W_1(\mu_1,\mu_2) = \min_{\mu} \sum_{i,j=1}^{n} d(x_i,x_j) \mu(x_i,x_j) \nonumber
\end{equation}
where $\mu(x,y)$ is a coupling (or flow) subject to the following constraints:
\begin{align}
\label{constraints}
\mu(x,y) &\geq 0 \hspace{9mm} \forall x,y \in X \\
\sum_{i=1}^{n} \mu(x, x_i) &= \mu_1(x) \hspace{3mm} \forall x\in X \\
\sum_{i=1}^{n} \mu(x_i,y) &= \mu_2(y) \hspace{3mm} \forall y \in X.
\end{align}
The cost above finds the optimal coupling of moving a set of mass from distributions $\mu_1$ to $\mu_2$ with minimal ``work''.

\subsection{Ricci Curvature and Mean-Reversion}
\begin{figure}[!t]
\begin{center}
\includegraphics[totalheight=4.5cm]{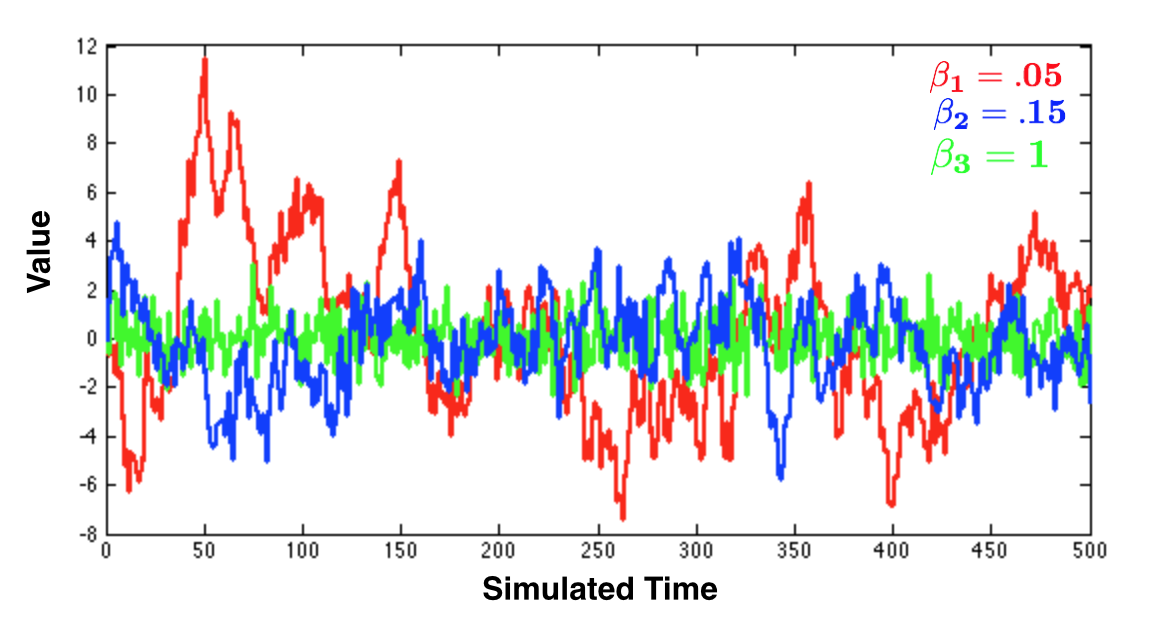}
\caption{We generated three Ornstein Uhlenbeck processes  $\beta_1=0.05$ (red), $\beta_2=0.15$ (blue), and $\beta_3 = 1.00$ (green) with computed corresponding Ollivier-Ricci curvatures to be $\kappa_1(x,y)=0.0488$,  $\kappa_1(x,y)=0.1393$, and $\kappa_1(x,y)=0.6321$ at $t=1$.  One can see that a signal that is perturbed with the same noise ($\sigma=1$), the signal with a higher curvature is capable of returning to equilibrium (zero) faster (e.g., more robust). }
\label{fig:OU_proc}
\end{center}
\end{figure}
Until now, we have connected Ricci curvature to fragility via entropy.  It is informative to further examine the behavior in the context of a decay rate when the system is perturbed.  As such, let us consider the relationship of Ollivier-Ricci curvature for an Ornstein-Uhlenbeck process.  More precisely, consider the stochastic differential equation
\begin{gather} \label{eq:OU}
dX_t = -\beta X_t dt + \sigma dW_t, \;\;  X(0) =x_0,
\end{gather}
where $W$ is the Wiener process (Brownian motion), and we take $x_0$ to be deterministic.  Then if we let $t>0$, it can be shown \cite{Oll_markov} that the Ollivier-Ricci curvature for this random walk for any two points $x_i$,$x_j\in \mathbb{R}^{n}$
\begin{equation} \label{ricci_ou}
\kappa(x_i,x_j) = 1- e^{-\beta t}.
\end{equation}
Equation~\ref{ricci_ou} illustrates the connection of fluctuation in a very simple explicit manner. Smaller $\beta$ corresponds to smaller curvature $\kappa$ and this corresponds to how slowly the systems returns to equilibrium, that is to the mean going to 0.  This can be seen in Figure \ref{fig:OU_proc}.  Lastly, while this is beyond the scope of this work, this may illuminate a new set of Ricci curvature based-strategies in statistical arbitrage and mean-reverting portfolios \cite{Aspremont}.
\begin{figure*}[!t]
\begin{center}
\includegraphics[totalheight=9.8cm]{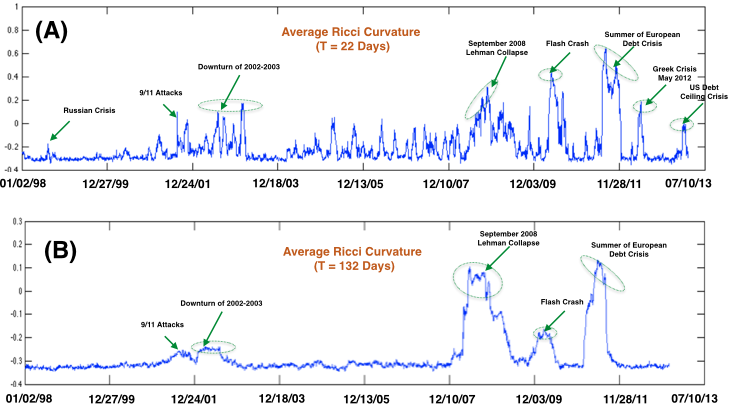}
\caption{We present the average Ollivier Ricci curvature of the 4000 networks generated from market return data on two different time scales:  (A) $T=22$ days and (B) $T=122$ days.  One can see that, during periods of known financial crisis, there is an increase in Ollivier-Ricci curvature compared to normal fragile market behavior.}
\label{fig:ricc_avg}
\end{center}
\end{figure*}

\subsection{Remarks on Leptokurtic Distributions}
Since the thrust of the present work is to understand systemic risk, we revisit the notion of leptokurtic distributions and their connection to Ricci curvature.
In particular, it has been argued that stock returns in the market should be modeled as heavy-tailed distributions as opposed to the standard normal distribution \cite{Rachev,RachevHT,mandel}.  This is done, in part, to better account for risk management in the event of financial crash and relatable black swan events.  This said, let us consider one such leptokurtic distribution, namely the Laplace distribution with a given mean $\theta$ and variance $\varphi$ and a corresponding normal distribution with the same mean and variance. Then, it is well known that entropy of the Laplacian $S^{\text{L}}_e$ and Gaussian $S^{\text{G}}_e$ distributions are given by the following:
\begin{align}
S_{e}^{\text{G}} &= \frac12 \log(2 \pi e \, \varphi^2)\\
S^{\text{L}}_e&=1+\log\bigg(2\sqrt{\frac{\varphi}{2}}\bigg).
\end{align}

Here, we see that $S^{\text{L}}$ increases more slowly than $S^{\text{G}},$ and thus a Laplacian process
is more fragile than the corresponding Gaussian process via the Fluctuation Theorem.  In short, the fat tail phenomenon used to model market returns is in fact attempting to account for an increase in market risk due to market fragility.  More interestingly, given that we have seen that $\Delta Ric \times \Delta S_e \geq 0$ (Ricci curvature is positively correlated with entropy), we may utilize Ricci curvature to analyze systemic risk as we will now experimentally show.

\section{Results}
\label{sec:results}

As a proof of concept, we present results on a set of stock correlation networks.  We again note, that the technique above can be applied to other interesting financial networks -- this will be a subject of new research.  We also note that the results here can viewed in the general sense (i.e., regardless of the actual network and construction thereof, there is a fascinating connection between network curvature and network entropy).

\subsection{Data, Network Generation}
We obtained historical closing daily price data from https://quantquote.com/.  In particular, the publicly available data set consists of stocks currently comprising the S\&P 500 for a 15 year span from January 1998 to July 2013.  We then filter those equities that do not have data for the entire period resulting in a total of 388 stocks.  From this, we are able to compute the correlation values $c_{xy}$ over a specific time-window denoted as $T$.  Then, following \cite{onnel1}, we construct a minimum spanning tree where the ``distance'' is defined to be $\hat{c}_{xy}: = \sqrt{2(1-c_{xy})}$.  This was done under the assumption that, at any given time, a particular stock must ``interact'' with another stock.  Furthermore, to examine the topology of the market through a dynamic process, we add high-valued links that satisfy a certain threshold, i.e., $c_{xy}\geq \xi$ where we chose $\xi=.85$  based on on the previous literature \cite{Bog,Chi} (e.g., this is due to the fact that forming a random walk over an MST will provide very similar structures).  This being said, for a given window $t^{b}=[t^{b}_0,...,t^{b}_{T}]$, we construct an unweighted network $\mathcal{N}^{b}$ through the above process.    A new network at $b+1$ is generated by ``sliding" the window of 1 day and repeating the process.  As such, approximately 4000 time-varying networks are generated.
\begin{figure}[!t]
\begin{center}
\includegraphics[totalheight=6cm]{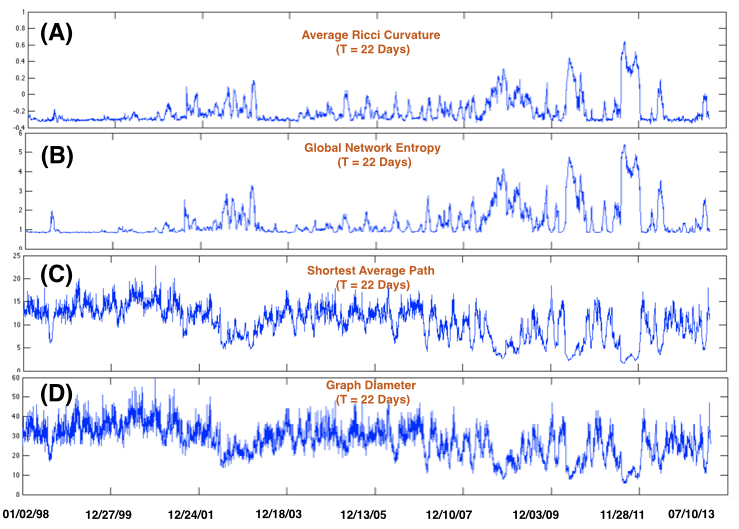}
\caption{We compare Ollivier-Ricci curvature (A) to that of network entropy (B) as well as shortest average path (C) and graph diameter (D) for a short time scale of 22 days.  There is a notable resemblance, as predicted, between network entropy and network curvature.  Further analysis shows that decreases in graph diameter and shortest path length result in increases in graph curvature.}
\label{fig:compare_1_month}
\end{center}
\end{figure}

\subsection{Market Fragility and Robustness of Crisis}
We begin by computing the Ollivier-Ricci curvature $\kappa(x,y)$ for all possible pairs ($\approx$ 75K pairs) over two different time windows $T=\{22, 132\}$ days.  As such, Figure \ref{fig:ricc_avg} presents the average Ricci curvature taken over the adjacency matrix such that it can be compared fairly with the information needed to compute global network entropy.  From analysis standpoint, revisiting equation (\ref{OR}), one can see curvature (and changes in curvature) for interactions ``far'' from the underlying topology will decay due to the term $d(x,y)$. Since a graph is a 1-geodesic space, if $\kappa(x,y)\geq k$ for $d(x,y)=1$, then $\kappa(x,y)\geq k$ $\forall$ $x,y$ \cite{Oll_markov}.  Thus, computing statistics (i.e., averages) for adjacent vertices will suffice and results are still valid (in the sense of robustness and fragility) for $d(x,y)\geq 2$, e.g., non-adjacent pairs in general will contribute negligibly and can be treated as scaling such statistic.

Interestingly, Figure~\ref{fig:ricc_avg} shows that the market operates in a generally fragile behavior.  As noted by prior works \cite{onnel1}, there is a topological market reorganization that occurs during periods of financial crisis.  This in part due to the establishment of more links in ``crash periods'' that naturally arises from the well known ``herding'' effect.  Constructing correlated graphs in a dynamic manner presents the market from a feedback perspective (by analyzing curvature) as opposed to merely an average correlation between any two given pairs of stocks.  In turn, Figure~\ref{fig:ricc_avg}
illustrates that even with two time scales, during periods of crisis there is an increase in Olliver-Ricci curvature and hence the market tends to become more robust.
This is consistent with previous analysis for which \cite{onnel1} examines the shortest average path between equities during such time periods and uncovered that the MST ``shrinks.''  Here, measuring Ollivier-Ricci curvature on these stock correlation networks illustrates fragile market behavior and the robustness of financial crashes.

\subsection{Comparison of Robustness Measures}
\begin{figure}[!t]
\begin{center}
\includegraphics[totalheight=7cm]{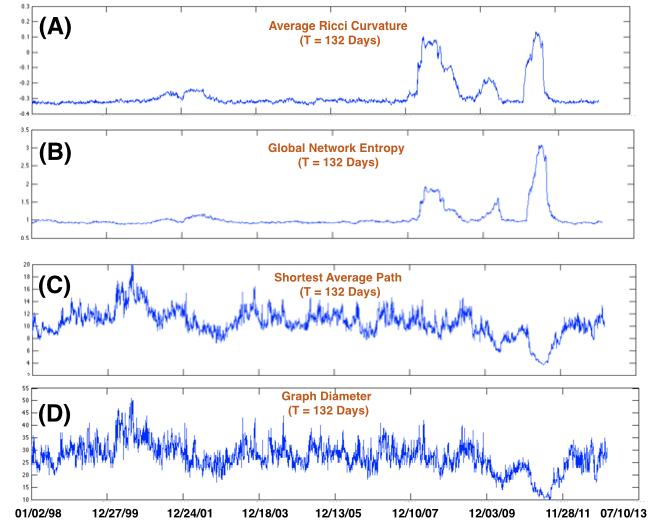}
\caption{We compare Ollivier-Ricci curvature (A) to that of network entropy (B) as well as shortest average path (C) and graph diameter (D) for a longer time scale of 132 days.  There is a notable resemblance, as predicted, between network entropy and network curvature.  Further analysis shows that decreases in graph diameter and shortest path length result in increases in graph curvature.}
\label{fig:compare_6_month}
\end{center}
\end{figure}

While the intended premise of this work is to focus on financial (correlation) networks with respect to curvature, it is significant to compare Ollivier-Ricci curvature to that of other well acceptable models characterizing network fragility.  To this end, Figure~\ref{fig:compare_1_month} and
Figure~\ref{fig:compare_6_month} plot curvature against that of global network entropy computed via equation (\ref{eq:network_entropy}),
shortest average path, and graph diameter.  One can see that there is a strikingly similar resemblance in the structure of Ollivier-Ricci curvature and global network entropy.  That is, we are uncovering the similar information to that of entropy with two very important caveats:  First, the method proposed herein is computationally more tractable and is well behaved compared to that entropy -- we are simply computing a linear programming (LP) problem which is capable of characterizing
curvature between any two nodes in a graph (not only those that are adjacent).  Secondly, Ollivier-Ricci curvature provides local geometric (edge) information about the network as opposed to entropy, which is a nodal measure.  In particular, one can visualize a scenario in a more complex financial network where a particular financial institution generally consists of normal risk exposures to other institutions with only a few extremely (indirect) risky transactions. By averaging such exposures, a nodal measure may not properly address localized fragility of such relationships.  This type of scenario has been shown by studying transcriptional networks
of varying cancer tissues where a single gene may participate in both robust and fragile interactions, yet is considered a ``robust'' gene \cite{cancer_arxiv}.  This being said, Figure~\ref{fig:compare_1_month} and Figure~\ref{fig:compare_6_month} also illustrate that increases in Ollivier-Ricci curvature are correlated
with a decrease in shortest average path and graph diameter.  Thus, Ollivier-Ricci curvature is an alternative method to expressing network functional robustness.
\begin{table*}[!ht]
\parbox{1\linewidth}{
\centering
\begin{tabular}{|c|c|c|c|c|c|c|c|c|c|c|c|c|c|c|c|c|}
\hline
\scriptsize{\textbf{Measure}} &   \scriptsize{Jan 98} &\footnotesize{Jan 99} & \scriptsize{Jan 00} & \scriptsize{Jan 01}  &\scriptsize{Jan 02} & \scriptsize{Jan 03} &\scriptsize{Jan 04} & \scriptsize{Jan 05} &\scriptsize{Jan 06} & \scriptsize{Jan 07}&  \scriptsize{Jan 08} &\scriptsize{Jan 09} & \scriptsize{Jan 10} &\scriptsize{Jan 11} & \scriptsize{Jan 12}\\
\hline
\scriptsize{\textbf{Curvature}}
&\scriptsize{-0.297}
& \scriptsize{-0.304}
& \scriptsize{-0.285}
& \scriptsize{-0.218}
& \scriptsize{-0.166}
& \scriptsize{-0.254}
& \scriptsize{-0.253}
& \scriptsize{-0.214}
& \scriptsize{-0.245}
& \scriptsize{-0.218}
& \scriptsize{-0.068}
& \scriptsize{-0.141}
& \scriptsize{-0.101}
& \scriptsize{0.023}
& \scriptsize{-0.255}\\
\hline
\scriptsize{\textbf{Entropy}}
&\scriptsize{0.941}
& \scriptsize{0.862}
& \scriptsize{0.909}
& \scriptsize{1.123}
& \scriptsize{1.505}
& \scriptsize{1.193}
& \scriptsize{1.022}
& \scriptsize{1.128}
& \scriptsize{1.066}
& \scriptsize{1.316}
& \scriptsize{2.159}
& \scriptsize{1.745}
& \scriptsize{2.105}
& \scriptsize{2.688}
& \scriptsize{1.282}	\\
\hline
\scriptsize{\textbf{Shortest Path}}
&\scriptsize{12.412}
& \scriptsize{13.811}
& \scriptsize{14.910}
& \scriptsize{12.951}
& \scriptsize{9.362}
& \scriptsize{10.199}
& \scriptsize{12.291}
& \scriptsize{11.845}
& \scriptsize{12.413}
& \scriptsize{9.636}
& \scriptsize{7.186}
& \scriptsize{8.060}
& \scriptsize{7.469}
& \scriptsize{6.208}
& \scriptsize{9.545}\\
\hline
\scriptsize{\textbf{Diameter}}
&\scriptsize{31.361}
& \scriptsize{34.726}
& \scriptsize{38.210}
& \scriptsize{33.186}
& \scriptsize{25.500}
& \scriptsize{27.397}
& \scriptsize{31.250}
& \scriptsize{29.929}
& \scriptsize{31.494}
& \scriptsize{25.064}
& \scriptsize{21.044}
& \scriptsize{22.480}
& \scriptsize{20.964}
& \scriptsize{17.770}
& \scriptsize{25.144}\\
\hline
\end{tabular}
\caption{We provide the average (average) Ricci curvature, average global network entropy, average (average) shortest path, and average graph diameter computed over 1-year period beginning with Jan 1st of each year with a window $T=22$ days and threshold of $\xi=.85$.  As seen in the previous graph, there exists a correlation between curvature to that of well known measures of fragility.}
\label{table:crap}}
\end{table*}

\subsection{Minimum Risk Markowitz Portfolios}
\label{sec:min_risk}

Let us now shift our focus to the role curvature and entropy (measures of fragility) play with respect to the classical Markowitz portfolio construction.  In the classic case, given a set of $N$ assets with average returns denoted as $\bar{r}=[\bar{r}_1,\bar{r}_2,...,\bar{r}_N]^{T}$ and the estimated covariance $\Sigma_r$ (both usually computed from historical data), one is given the task allocating $N$ asset weights $w=[w_1,w_2,....,w_N]^{T}$ in order to maximize the portfolio return $u = \sum_i^n = w^T \bar{r}$ subject to some risk tolerance $w^{T}\Sigma w \leq \tau_{\text{risk}}$ with $\sum_i^N w = 1$ and $w_i\geq 0$ (no short selling).  The efficient frontier portfolios \cite{marko} then can be computed for varying levels of risk tolerance -- here we are interested in the minimum risk portfolio and connections to curvature.

\begin{figure}[h]
\begin{center}
\includegraphics[totalheight=4.5cm]{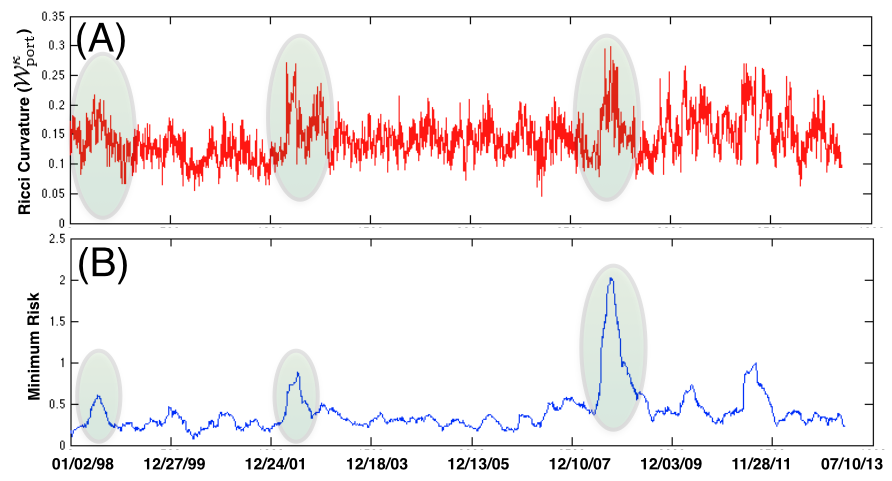}
\caption{We compute $\mathcal{W}^{\kappa}_{\text{port}}$ using weights from the minimum risk portfolio along with minimum risk.  Regions of interested are highlighted for (A) Ricci curvature $\mathcal{W}^{\kappa}_{\text{port}}$ and (b) minimum risk that can be obtained along the efficient frontier over a given window - $T=122$ days with a threshold $\xi=0.85$ was chosen. }
\label{fig:min_risk}
\end{center}
\end{figure}

It is known that a decrease in entropy relates to an increase in diversification with several attempts to capture this property in portfolio construction (see \cite{entropy_finance} and references therein).  From the perspective in the present work, one see that there is a lack of feedback when constructing such correlation networks in normal market behavior (decrease in curvature) and this may be a desirable property when seeking out a diversified portfolio (e.g., those pairs that exhibit fragile interactions are candidates for diversification). With this in mind, we propose the following measure which is simply a projection of Markowitz portfolio weights from the minimum risk portfolio onto edge Ollivier-Ricci curvature:
\begin{equation}
\mathcal{W}^{\kappa}_{\text{port}} = \sum_i^n w^{T}\kappa w
\end{equation}
We note that the above measure is dependent on a particular graph construction.  That said, Figure~\ref{fig:min_risk} shows $\mathcal{W}^{\kappa}_{\text{port}}$ along with the minimum risk profile.  One can see that there exists a correlation between increase in curvature and increase in minimum risk profile.
Moreover, the above trend seemingly falls in line with previous analysis -- during periods of crisis, it has already been noted that diversification ``melts away'' \cite{preis}.  This is analogous to increases in curvature that is seen in Figure~\ref{fig:ricc_avg} and
Figure~\ref{fig:min_risk}.  On the other hand, we have noted that higher Ollivier-Ricci curvature is positively correlated to the mean reverting coefficient
($\Delta Ric \times \Delta \beta$) in an Ornstein-Uhlenbeck sense.  Thus, increases in Ricci curvature from a portfolio construction relates to the notion of possibly forming a ``risk-free'' market neutral portfolio (e.g., a perturbation in the portfolio will quickly return to some equilibrium).  \textbf{We caution the reader on the above suggestive points as it will be subject to future work and requires a much further (and more thoughtful) analysis.}

\section{Summary \& Conclusion}
\label{sec:summary}
Given the 2007-2008 financial crisis and European debt crisis, there is an increasing importance of understanding fragility of interconnected networks.  This work proposes a novel new metric based on a coarse definition of Ricci curvature in order to understand systemic risk of a complex web of interacting agents.  Due to initial data limitations, we provided a proof of concept in the context of stock correlation networks to elucidate market fragility.  This being said, there seem to very interesting connections of Ricci curvature to not only financial stability of varying networks, but also in the construction of market neutral strategies and possibly novel risk management policies.  In particular, one can begin to explore the discrete Ollivier-Ricci flow (i.e., $\frac{d}{dt} d(x,y)=-\kappa(x,y)d(x,y)$) in order to combat and prevent financial contagions - this has been suggested in related fields \cite{Jonck} to remove overload queues in wireless networks.  One of the main subjects of future of research is to continue to explore and analyze systemic risk in more interesting financial networks where the rise of complex financial instruments \cite{bat3} may prove to be a great source of financial instability.

Lastly, it is worthwhile to note here that these results hold in general and can be very much applicable in studying a variety of dynamic networks where high frequency data is available, i.e., high-throughput genomic studies characterizing the progression of metastatic cancer is one such area that is readably applicable.   Whereas we seek drug targets to combat the robust nature of cancer \cite{cancer_arxiv}, we are at the same time seeking to combat the fragile nature of market from financial contagions.  We further note some interesting work by Billera \cite{Billera} who describes a metric geometry on the space of trees in connection with phylogenetics with a space that possess non-positive curvature.  From the results of \cite{Sturm},  this allows one to do statistics on this space since between any two points there is a unique geodesic.  Here, we can then generalize the analysis from a single network structure to devising statistical methods comparing families of networks, e.g., market returns, genomic studies, etc. \cite{Rabadan}.

In summary and following our previous results \cite{cancer_arxiv}, \textbf{\emph{we want to establish such policies that allow the market to be more cancer like}} by providing novel ``drug targets'' to contain financial contagions such as the 2007-2008 crisis that nearly brought down the financial system.

\end{document}